# Securing the AI Frontier: Urgent Ethical and Regulatory Imperatives for AI-Driven Cybersecurity

Vikram Kulothungan
*Capitol Technology University*
North Bergen, U.S.A
vikramk1986@gmail.com

*Abstract*—This paper critically examines the evolving ethical and regulatory challenges posed by the integration of artificial intelligence (AI) in cybersecurity. We trace the historical development of AI regulation, highlighting major milestones from theoretical discussions in the 1940s to the implementation of recent global frameworks such as the European Union's AI Act. The current regulatory landscape is analyzed, emphasizing risk-based approaches, sector-specific regulations, and the tension between fostering innovation and mitigating risks. Ethical concerns—such as bias, transparency, accountability, privacy, and human oversight—are explored in depth, along with their implications for AI-driven cybersecurity systems. Furthermore, we propose strategies for promoting AI literacy and public engagement, essential for shaping a future regulatory framework. Our findings underscore the need for a unified, globally harmonized regulatory approach that addresses the unique risks of AI in cybersecurity. We conclude by identifying future research opportunities and recommending pathways for collaboration between policymakers, industry leaders, and researchers to ensure the responsible deployment of AI technologies in cybersecurity.

## I. Introduction

Artificial Intelligence (AI) has emerged as a transformative force in cybersecurity, offering unparalleled capabilities in threat detection, incident response, and risk management. With its ability to process vast amounts of data in real-time, AI systems can identify cyber threats with unprecedented speed and accuracy, predicting vulnerabilities and enhancing overall security [1]. However, the rapid integration of AI into cybersecurity also raises significant ethical and regulatory concerns. These concerns include privacy violations, bias in AI-driven decision-making, lack of transparency, and diminished human oversight [2].

As AI technologies advance, the absence of robust regulatory frameworks and ethical guidelines has become increasingly problematic. The global regulatory landscape remains fragmented, with different countries and industries adopting varying approaches to AI governance. The European Union's AI Act [3], the U.S. Executive Order on AI [4], and other sector-specific initiatives illustrate attempts to regulate AI, yet these efforts often fall short of addressing the complex, cross-border nature of AI and cybersecurity.

This paper seeks to address these challenges by providing a comprehensive review of the ethical and regulatory issues surrounding AI in cybersecurity. We examine the historical evolution of AI regulation, highlight key ethical considerations, and explore current regulatory frameworks. By analyzing the interplay between innovation and risk, we aim to offer insights into the development of globally harmonized regulatory mechanisms.

The scope of this paper includes:

- A detailed examination of AI regulation from its early theoretical roots to modern-day frameworks.
- An analysis of current regulatory approaches, focusing on risk-based and sector-specific regulations.
- A discussion of key ethical challenges, including fairness, transparency, accountability, privacy, and human oversight in AI-driven cybersecurity systems.
- Recommendations for future research directions for AI governance in cybersecurity.

By addressing these areas, this paper aims to contribute to ongoing discussions and provide actionable insights for researchers, policymakers, and industry leaders working at the intersection of AI, cybersecurity, and regulation.

## II. Historical Evolution of AI Regulation

The regulation of artificial intelligence (AI) has evolved significantly over the past several decades, reflecting the increasing sophistication of AI technologies and growing awareness of their societal impacts. This evolution can be divided into four key phases: early awareness, the emergence of ethical guidelines, the development of initial regulatory frameworks, and the current acceleration toward global governance. Understanding this historical progression is essential for identifying both the achievements and shortcomings of AI regulation today.

### A. Early Awareness(1940s-early 2000s)

The foundations of AI regulation were laid during the mid-20th century, though discussions remained largely theoretical at this stage. Alan Turing's seminal 1950 paper, "Computing Machinery and Intelligence," introduced the idea of machine intelligence, igniting debates about the future implications of AI technologies [5]. Similarly, the Dartmouth Conference of 1956 marked the formal birth of AI as a field, though it was decades before serious consideration was given to regulatory concerns [6].

Throughout the late 20th century, discussions about AI regulation centered on existential risks and long-term ethical dilemmas, with little concrete action taken. Academic institutions such as the Future of Humanity Institute (founded in 2005) began to address the potential dangers posed by advanced AI systems [7]. However, regulatory efforts remained minimal during this period, as AI's practical applications in fields like cybersecurity had not yet fully materialized.

## B. Emergence of Ethical Guidelines (2010-2015)

As AI technologies became more prevalent in the early 2010s, the need for ethical guidelines became evident. Concerns about bias, fairness, transparency, and accountability began to surface as AI systems started being deployed in real-world applications, including cybersecurity.

In 2014, the European Parliament passed one of the first legislative efforts to address AI, adopting a resolution on "Civil Law Rules on Robotics" [8]. This was followed by the U.S. government's 2016 report, "Preparing for the Future of Artificial Intelligence," which called for proactive measures to ensure AI safety and ethical use [9]. Around the same time, the Institute of Electrical and Electronics Engineers (IEEE) launched its "Ethically Aligned Design" initiative, which laid out principles for the ethical development of autonomous systems [10]. These early guidelines, while not legally binding, represented a growing recognition of the need to manage AI's risks responsibly.

## C. Initial Regulatory Frameworks (2016-2020)

By the mid-2010s, AI technologies had advanced to a point where ethical guidelines were no longer sufficient on their own, prompting the development of formal regulatory frameworks. These efforts reflected a shift from theoretical discussions to concrete action aimed at managing AI's growing role in society.

In 2016, the Partnership on AI was formed by major technology companies to promote best practices in AI development and governance [11]. Meanwhile, the European Commission's High-Level Expert Group on AI published the "Ethics Guidelines for Trustworthy AI" in 2019, which provided detailed recommendations on ensuring that AI systems are lawful, ethical, and robust [12]. This period also saw the Organization for Economic Co-operation and Development (OECD) adopt the "Principles on AI" in 2019, a landmark agreement endorsed by 42 countries to establish international standards for AI governance [13].

In the U.S., the government issued its "Guidance for Regulation of Artificial Intelligence Applications" in 2020, signaling a clear effort to regulate AI in sectors such as cybersecurity, emphasizing transparency, fairness, and public trust [14].

## D. Acceleration and Global Focus (2021-Present)

In recent years, the pace of AI regulation has accelerated dramatically, driven by the rapid development of advanced AI technologies, such as large language models and generative AI. The global impact of AI on industries like cybersecurity has underscored the need for comprehensive and adaptable governance frameworks.

The European Union has led these efforts with the introduction of its AI Act, which is set to take effect in 2024. This groundbreaking legislation adopts a risk-based approach, categorizing AI systems based on their potential impact and imposing proportionate regulatory measures [3]. Similarly, the United States has intensified its regulatory focus, with the issuance of the 2023 Executive Order on "Safe, Secure, and Trustworthy Development and Use of AI," which emphasizes safety and security standards for AI applications [15].

International cooperation has also gained momentum, as seen in UNESCO's 2021 "Recommendation on the Ethics of Artificial Intelligence," which sets a global benchmark for ethical AI development [16]. The expansion of AI-related bills in national legislatures—from 88 in 2022 to 181 in 2023 in the U.S. alone—illustrates the growing recognition of AI's societal impact "Fig. 1" [17].

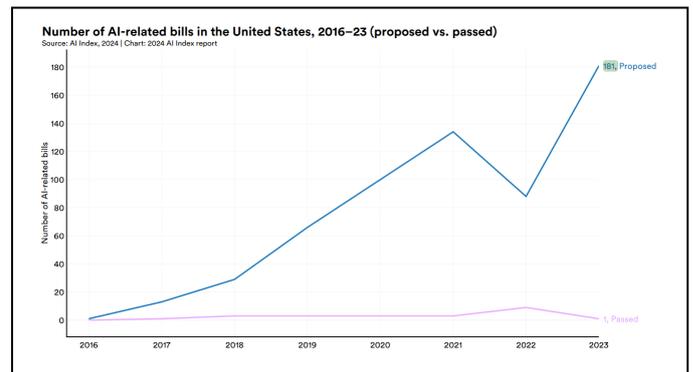

Fig. 1. Number of AI-related bills in the United States, 2016-23. (*proposed vs passed*)

## E. Lessons Learned and Path Forward

The historical evolution of AI regulation reveals both progress and persistent challenges. While early ethical guidelines laid a critical foundation, they lacked enforcement mechanisms, resulting in regulatory gaps that continue to challenge policymakers. Initial regulatory frameworks, though important, often fail to keep pace with the rapid technological advancements of AI. The current phase of global focus and risk-based regulation marks a significant step forward, but ongoing adaptation and harmonization are essential to manage AI's cross-border impacts effectively.

As AI technologies continue to evolve, particularly in critical sectors like cybersecurity, regulatory frameworks must become more flexible and globally coordinated. Future efforts should focus on creating "living" regulatory frameworks that can adapt in real-time to emerging ethical concerns and technological advancements.

## III. CURRENT REGULATORY LANDSCAPE

The current regulatory landscape for artificial intelligence (AI) in cybersecurity is marked by diverse approaches, reflecting the complexity and rapid evolution of AI technologies. While governments and international organizations have made significant progress in developing frameworks to govern AI, challenges remain in achieving harmonization and balancing innovation with risk mitigation. This section outlines the key trends in AI regulation, focusing on risk-based frameworks, sector-specific regulations, innovation-risk trade-offs, and global harmonization efforts.

## A. Risk-Based Frameworks

A prominent trend in AI regulation is the adoption of risk-based frameworks, which classify AI systems according to their potential impact and apply proportionate regulatory measures. This approach acknowledges that not all AI applications pose the same level of risk to individuals or society, allowing for more targeted oversight where necessary.

The European Union's AI Act, set to be fully implemented in 2024, is one of the most comprehensive examples of a risk-based regulatory model. The act categorizes AI systems into four risk levels: unacceptable risk, high risk, limited risk, and minimal risk [3]. For AI systems deployed in critical infrastructure, including cybersecurity applications, the "high risk" designation mandates stringent compliance requirements. These include strict oversight, data quality standards, transparency measures, and human oversight protocols.

This risk-based approach is gaining traction globally due to its flexibility. It allows regulators to focus their efforts on the most potentially harmful AI systems while enabling less burdensome regulations for lower-risk applications. However, challenges remain in defining and enforcing these risk categories consistently across jurisdictions, especially as AI technologies and cybersecurity threats evolve rapidly.

*B. Sector-Specific Regulation*

While general AI regulatory frameworks provide a broad governance structure, many sectors—including cybersecurity—require tailored regulations to address specific challenges. In this context, several countries and regions have introduced sector-specific regulations that complement broader AI governance frameworks.

In the financial services sector, which is closely tied to cybersecurity, the U.S. Consumer Financial Protection Bureau (CFPB) has issued guidance on the use of AI in credit decision-making. This guidance emphasizes fairness, accountability, and explainability, principles that have direct implications for cybersecurity measures within the financial industry [18]. Similarly, the healthcare sector, another field with significant cybersecurity risks, has developed regulations focused on the ethical and secure use of AI in medical devices and patient data protection. These regulations often prioritize data privacy, underscoring the importance of security measures when handling sensitive information.

In cybersecurity, where real-time decision-making is crucial, sector-specific regulations often emphasize the need for rapid response mechanisms and robust incident reporting frameworks. As AI becomes more integral to cybersecurity operations, particularly in threat detection and mitigation, sector-specific guidelines will likely expand to cover the ethical use of AI in sensitive applications.

*C. Balancing Innovation with Risk Mitigation*

One of the greatest challenges in AI regulation is balancing the need to foster innovation with the imperative to mitigate risks. Overly stringent regulations can stifle technological advancement, particularly in fast-moving fields like AI-driven cybersecurity, where new threats and solutions emerge constantly. Conversely, lax regulations can lead to the unchecked deployment of AI systems, increasing the likelihood of security breaches, bias, and other negative outcomes.

To address this tension, several regulatory models have emerged that aim to promote innovation while ensuring sufficient oversight. Regulatory sandboxes, for example, have gained popularity as controlled environments where companies can test AI applications without being subjected to full regulatory requirements. This allows for experimentation and innovation while maintaining regulatory oversight [19]. Sandboxes have been particularly useful in cybersecurity, enabling companies to develop AI-driven threat detection and response tools that comply with core ethical and security standards without being hindered by excessive regulation.

The concept of "agile governance" has also gained traction, particularly in countries like Japan, where regulatory frameworks are designed to be flexible and responsive to technological advancements. This approach ensures that regulations can be rapidly updated as AI technologies evolve, which is especially important in the cybersecurity domain, where the stakes of regulatory delays can be severe.

*D. Global Harmonization Efforts*

Key global and regional regulatory frameworks can be compared based on their scope, implementation timelines, and focus areas. "Table I." below summarizes the approaches adopted by the EU, the U.S., and OECD, illustrating critical distinctions in their methodologies.

TABLE I. EU AI ACT ALIGNMENT

| Framework | Scope | Focus Areas |
|---|---|---|
| EU AI Act (Europe) | Risk-Based Framework | Any Developers and Deployers of AI systems in Public |
| U.S Executive Order (United States) | Decentralized and sectoral | Transparency, Safety, and Innovation |
| OECD Principles (International) [13] | Voluntary guidelines | Ethics, human-centered design |

Fragmented regulatory approaches create challenges for multinational organizations and can lead to loopholes in governance, where AI systems that comply with regulations in one country may pose risks in another. Achieving greater alignment between these frameworks will require ongoing dialogue, international collaboration, and perhaps the development of a global AI cybersecurity consortium.

*E. Challenges in Implementation*

While the development of AI regulatory frameworks represents significant progress, their implementation poses a range of challenges, particularly in the context of cybersecurity. The rapid pace of technological advancement often outstrips the speed of regulatory processes, resulting in potential governance gaps. Moreover, AI systems used in cybersecurity are often highly complex, making it difficult to apply traditional regulatory approaches, particularly regarding algorithmic transparency and accountability.

Another key challenge is jurisdictional: cybersecurity threats and AI systems frequently transcend national borders, complicating efforts to enforce regulations across different legal contexts. Even as international cooperation intensifies, regulatory disparities between nations can lead to gaps in global cybersecurity protection, allowing bad actors to exploit weaker regulatory environments. Small and medium enterprises (SMEs) struggle with the financial burden of aligning with harmonized regulatory standards, particularly in high-risk sectors like cybersecurity. Regulatory sandboxes can help test AI applications across jurisdictions, fostering innovation while ensuring compliance. Establishing a global AI cybersecurity consortium would also facilitate cross-border cooperation and harmonization.

IV. Ethical Considerations in AI Deployment for Cybersecurity

The integration of artificial intelligence (AI) into cybersecurity operations presents a range of ethical challenges that must be carefully addressed to ensure responsible and equitable deployment. As AI becomes increasingly autonomous in identifying threats and responding to cyber incidents, questions of fairness, transparency, accountability, privacy, and human oversight become paramount. This section explores these key ethical considerations and their implications for AI-powered cybersecurity systems.

*A. Fairness and Non-Discrimination*

One of the most pressing ethical concerns in AI deployment is the potential for bias and discrimination, particularly in decision-making processes that can affect individuals or groups. AI systems trained on historical data may unintentionally perpetuate or exacerbate existing biases, leading to unfair outcomes in cybersecurity contexts. For instance, certain demographic groups could be disproportionately flagged as security risks due to biased training data, resulting in unwarranted scrutiny or denial of services.

In cybersecurity, bias in AI algorithms could manifest in user profiling, risk scoring, or threat detection, where certain individuals or communities are unfairly targeted based on erroneous or biased data inputs [2]. Ensuring fairness in AI-powered cybersecurity systems requires careful attention to the quality and representativeness of training data, as well as ongoing monitoring to detect and mitigate bias in real-time.

Strategies for promoting fairness include:

- Diverse and representative datasets: AI models should be trained on datasets that reflect a wide range of user behaviors, demographics, and contexts to avoid reinforcing existing biases.

- Bias detection and mitigation: Implement continuous auditing mechanisms to identify and address biased outcomes in cybersecurity operations, ensuring that AI systems do not disproportionately affect vulnerable populations.

*B. Transparency and Explainability*

The "black box" nature of many AI systems, particularly deep learning models, poses significant challenges for transparency and explainability in cybersecurity applications. In cybersecurity, AI often makes complex, high-stakes decisions—such as identifying threats or determining the appropriate response to an incident—yet the rationale behind these decisions is often opaque to human operators. This lack of transparency can erode trust in AI systems, making it difficult for security teams to justify or understand AI-driven actions.

Explainable AI (XAI) is critical in cybersecurity contexts, where decision-makers need clear explanations of why a particular threat was detected or how an AI system arrived at a specific conclusion [20]. This transparency is not only essential for operational effectiveness but also for legal and ethical accountability.

Challenges in achieving transparency include:

- Complexity of AI models: Advanced AI models, such as neural networks, are inherently difficult to interpret, making it challenging to provide clear explanations without oversimplifying critical information.

- Trade-offs between performance and explainability: Increasing transparency may sometimes reduce the performance of AI systems, particularly in time-sensitive cybersecurity applications where speed and accuracy are paramount.

Approaches to enhancing explainability:

- Hybrid models: Combining deep learning with rule-based systems or simpler, interpretable models can help bridge the gap between performance and transparency.

- Post-hoc explanations: Techniques such as SHAP (SHapley Additive exPlanations) or LIME (Local Interpretable Model-agnostic Explanations) can provide insights into how AI models make decisions, even after predictions have been made.

*C. Accountability and Liability*

As AI systems become more autonomous in managing cybersecurity threats, the question of accountability becomes increasingly complex. Determining responsibility for AI-related actions—particularly in the case of security breaches or false positives—poses significant challenges [21]. When an AI system incorrectly flags a legitimate action as malicious or fails to detect a genuine threat, it can be difficult to ascertain whether accountability lies with the AI developers, the cybersecurity teams deploying the system, or the organizations using it.

The issue of liability is further complicated by the dynamic and evolving nature of cyber threats. AI systems, which are trained to detect certain patterns based on historical data, may struggle to adapt to novel threats or changing environments without human oversight, raising questions about the extent to which humans should remain "in the loop."

Key issues in accountability include:

- Liability for AI errors: Legal frameworks need to clearly define the responsibility of AI developers and end-users in cases where AI-driven decisions result in harm or security breaches.

- Human oversight: Ensuring that AI systems are subject to meaningful human oversight is crucial for maintaining accountability in high-stakes cybersecurity operations.

*D. Privacy and Data Protection*

AI systems deployed in cybersecurity often require access to vast amounts of data to function effectively, raising significant concerns about privacy and data protection. The need to analyze extensive data streams—including sensitive personal information—can lead to privacy violations, particularly if AI systems are not designed with privacy in mind. Moreover, cybersecurity AI systems may inadvertently collect more data than necessary or fail to anonymize it properly, posing additional risks to users' privacy rights.

Balancing the need for data in AI-driven threat detection with privacy concerns is a delicate task [22]. Ethical AI deployment in cybersecurity must adhere to privacy regulations such as the General Data Protection Regulation (GDPR) and ensure that data collection and processing are both necessary and proportional to the threat being addressed.

Ethical considerations in privacy include:

- Data minimization: AI systems should be designed to collect only the data necessary for cybersecurity purposes, reducing the risk of privacy violations.

- Privacy-preserving techniques: Techniques such as federated learning, which allows AI systems to train across multiple datasets without sharing sensitive data, and homomorphic encryption, which enables the analysis of encrypted data, offer promising ways to protect privacy while maintaining security.

*E. Human Oversight and Control*

Maintaining appropriate human oversight in AI-driven cybersecurity systems is critical for ensuring that these systems operate ethically and effectively. As AI systems take on more autonomous roles in detecting and responding to threats, it is essential to preserve human control, particularly in high-stakes scenarios where mistakes can have severe consequences. The principle of "human-in-the-loop" governance—where humans retain final decision-making authority over AI actions—is especially important in cybersecurity, where false positives or negatives can lead to significant operational disruptions or vulnerabilities [1].

Challenges in implementing human oversight include:

- Over-reliance on automation: As AI systems become more advanced, there is a risk that human operators may become overly dependent on AI-driven decisions, reducing their ability to intervene when necessary.

- Defining clear roles: Establishing well-defined roles and responsibilities for human operators in AI-augmented cybersecurity environments is crucial for ensuring that human oversight is meaningful and effective.

Best practices for human oversight:

- Tiered decision-making: AI systems can be programmed to escalate high-risk or uncertain decisions to human operators for review, ensuring that critical decisions are subject to human scrutiny.

- Continuous training: Ongoing education and training programs for cybersecurity professionals can help them effectively collaborate with AI systems and intervene when necessary.

*F. Societal Impact and Workforce Considerations*

The deployment of AI in cybersecurity not only raises technical and ethical questions but also has significant implications for society and the workforce. As AI systems take on increasingly complex roles in threat detection and cybersecurity management, they are likely to reshape the workforce, alter skill requirements, and raise concerns about job displacement and societal equity.

*1) Job Displacement and Reskilling:*

One of the most significant societal impacts of AI in cybersecurity is the potential displacement of jobs traditionally performed by human analysts. AI systems can process vast amounts of data, identify patterns, and make real-time decisions far more efficiently than human operators. As these systems become more autonomous, there is concern that some cybersecurity roles may become redundant, particularly in areas like routine monitoring and basic threat detection.

However, while AI may replace certain jobs, it is also expected to create new roles that require advanced technical expertise. The increasing complexity of AI systems means that organizations will need highly skilled professionals who can manage, interpret, and maintain AI tools. This shift from manual to automated processes will demand significant investment in reskilling and upskilling programs to prepare the current workforce for AI-enhanced roles.

Key workforce considerations include:

- Job displacement: Routine cybersecurity tasks, such as log analysis and initial threat detection, may be automated, potentially reducing demand for entry-level analysts.

- Reskilling and upskilling: AI will create demand for new skills, such as data science, AI system management, and ethical AI governance, requiring cybersecurity professionals to adapt and expand their expertise.

- Human-AI collaboration: Rather than completely replacing human workers, AI systems will augment human decision-making, particularly in complex and high-stakes situations. Training employees to effectively collaborate with AI systems will be critical to maximizing the benefits of these technologies.

*2) Equity and Access:*

As AI technologies become more integral to cybersecurity, issues of equity and access will need to be addressed. There is a risk that organizations with greater resources and technical expertise may benefit disproportionately from AI advancements, widening the gap between well-funded institutions and smaller entities that lack the ability to implement cutting-edge AI solutions. This disparity could lead to unequal protection levels across different sectors and regions, particularly in critical infrastructure and public services.

Moreover, the use of AI in cybersecurity could exacerbate existing societal inequalities if certain groups are unfairly targeted or excluded from digital protections due to biased algorithms or limited access to advanced cybersecurity technologies. For instance, under-resourced communities or developing nations may find it more difficult to deploy AI-driven cybersecurity measures, leaving them more vulnerable to cyber threats.

Ethical implications of equity include:

- Digital divide: Smaller organizations, governments, or regions may struggle to adopt advanced AI technologies due to financial or technical constraints, leading to uneven protection against cyber threats.

- Bias in access: AI systems that rely on biased datasets may inadvertently exclude or over-target specific demographic groups, creating a digital divide in terms of cybersecurity protections.

*3) Workforce Diversity in AI and Cybersecurity:*

Addressing the ethical challenges of AI in cybersecurity also requires attention to workforce diversity. Ensuring a diverse set of perspectives in the development and deployment of AI systems is essential for reducing bias and improving the overall fairness of AI solutions. The current underrepresentation of women and minority groups in both AI and cybersecurity fields contributes to the risk of biased decision-making and exclusionary practices in AI system design.

Strategies for promoting diversity include:

- Inclusive hiring practices: Organizations should actively work to build diverse teams of AI developers, cybersecurity professionals, and policymakers to ensure that AI systems are designed with varied perspectives in mind.
- Education and outreach: Expanding access to AI and cybersecurity education programs, particularly for underrepresented groups, can help address disparities in the workforce and improve the inclusivity of AI-driven cybersecurity tools.

## V. FUTURE DIRECTIONS AND RESEARCH OPPORTUNITIES

As artificial intelligence (AI) continues to advance, its applications in cybersecurity will evolve, presenting both new opportunities and challenges. To ensure the responsible and effective deployment of AI in cybersecurity, future research must address critical areas where the current regulatory, ethical, and technological frameworks are either lacking or need enhancement. This section identifies key areas for future research and policy development, emphasizing the need for adaptive regulatory frameworks, quantum computing preparedness, ethical decision-making, cross-border collaboration, and improved transparency.

### A. Adaptive Regulatory Frameworks

Given the rapid pace of AI advancements, static regulatory frameworks risk becoming outdated before they can effectively govern new technologies. To address this, future research should focus on developing adaptive, "living" regulatory frameworks that can evolve in real-time in response to emerging AI trends and threats. These frameworks should be capable of continuous updates based on new technological developments, ethical concerns, and cybersecurity risks.

*1) Research Opportunity:*

Real-time regulatory updates: Investigate methodologies for creating AI-assisted regulatory systems that can analyze ongoing trends in AI development and propose updates to governance mechanisms accordingly. Such systems could provide early warnings about emerging risks or vulnerabilities and suggest timely regulatory changes.

*2) Future Direction:*

AI-enhanced regulatory tools: Explore the potential of using AI to assist in real-time regulatory enforcement, enabling continuous monitoring and adaptation of cybersecurity practices to align with rapidly evolving technologies.

### B. Quantum Computing and AI Security

The advent of quantum computing poses both significant challenges and opportunities for AI in cybersecurity. Quantum computers could potentially render current cryptographic standards obsolete, opening new vulnerabilities for AI-powered cybersecurity systems. At the same time, quantum computing could enhance AI's ability to solve complex cybersecurity problems, such as threat modeling and encryption.

*1) Research Opportunity:*

Quantum-resistant AI algorithms: Research into quantum-resistant algorithms is crucial to ensure that AI-powered cybersecurity systems can withstand future quantum threats. This includes developing encryption protocols that remain secure in the post-quantum era.

Also, Quantum computing has the potential to enhance AI in solving complex cybersecurity challenges. For instance, research into quantum algorithms that could optimize threat modeling and improve large-scale encryption protocols.

*2) Future Direction:*

Quantum-safe AI systems: Develop AI architectures that incorporate quantum-resistant cryptographic methods, ensuring that AI systems remain effective and secure even as quantum computing capabilities advance.

### C. Ethical AI Decision-Making in Cybersecurity

As AI systems become more autonomous in their decision-making processes, particularly in high-stakes cybersecurity environments, ensuring that these decisions are ethically sound becomes increasingly important. Embedding ethical considerations directly into AI algorithms can help prevent biased, unfair, or harmful outcomes.

*1) Research Opportunity:*

Embedding ethics in AI algorithms: Develop frameworks for integrating ethical decision-making principles into AI algorithms used in cybersecurity. This includes ensuring that AI systems prioritize fairness, transparency, and accountability in threat detection and response scenarios.

*2) Future Direction:*

Ethical "black boxes": Explore the creation of ethical "black boxes" for AI systems in cybersecurity, similar to flight data recorders in aircraft. These tools would enable post-hoc analysis of AI decisions, helping to assess the ethical implications of AI actions and improve accountability.

### D. Cross-Border Colloboration and Global Standards

Cyber threats transcend national borders, making international cooperation essential for effective AI regulation and cybersecurity governance. The development of global standards for AI in cybersecurity can help ensure consistent protection levels across regions and prevent regulatory fragmentation that could be exploited by bad actors.

*1) Research Opportunity:*

Global governance frameworks: Analyze the effectiveness of existing international collaborations, such as the OECD AI Principles and UNESCO's ethics recommendations, to identify best practices for global cooperation in AI regulation. Research should also explore the creation of new, more comprehensive frameworks for international AI governance in cybersecurity.

*2) Future Direction:*

International AI-cybersecurity consortium: Propose the development of a global AI-cybersecurity consortium that facilitates cross-border collaboration between governments, industry leaders, and researchers. Such a consortium could help harmonize regulatory approaches, share intelligence, and address jurisdictional challenges in global cybersecurity threats.

### E. AI Transparency and Explainability in Cybersecurity

Improving the transparency and explainability of AI systems in cybersecurity is crucial for building trust, ensuring accountability, and enhancing operational effectiveness. As AI systems make more autonomous decisions in cybersecurity contexts, stakeholders need to understand and validate those decisions, particularly in high-risk scenarios.

*1) Research Opportunity:*

Explainability techniques: Develop new methods for making complex AI models more interpretable without compromising their effectiveness. This includes research into explainability techniques tailored specifically to cybersecurity applications, where the need for speed and accuracy often conflicts with the desire for transparency.

*2) Future Direction:*

Standardized explainability metrics: Explore the potential for developing standardized explainability metrics for AI systems in cybersecurity. These metrics could help organizations assess how transparent and interpretable their AI models are, providing benchmarks for AI-driven cybersecurity tools.

*F. Human-AI Colloboration in Cybersecurity*

Optimizing the interaction between human analysts and AI systems is critical for improving cybersecurity outcomes. While AI systems can process large datasets and detect patterns that humans might miss, human expertise is still essential for interpreting ambiguous situations, making final decisions, and responding to novel or complex threats.

*1) Research Opportunity:*

Cognitive models of human-AI collaboration: Investigate cognitive models that can optimize the division of labor between human analysts and AI systems in cybersecurity. Research should focus on identifying which tasks are best handled by AI and which require human intervention, as well as improving the interfaces through which humans interact with AI systems.

*2) Future Direction:*

Adaptive AI interfaces: Develop adaptive interfaces that adjust the level of AI autonomy based on the expertise and cognitive load of human operators. These interfaces should provide varying degrees of control and explainability, depending on the complexity of the task and the human operator's experience.

*G. AI Literacy and Cybersecurity Education*

As AI continues to transform cybersecurity, there is a growing need to enhance AI literacy among cybersecurity professionals and the general public. Effective AI literacy programs can empower users to engage critically with AI technologies, better understand their risks and benefits, and contribute to the development of ethical and effective AI systems.

*1) Research Opportunity:*

Evaluating AI-cybersecurity education programs: Conduct systematic reviews of current AI-cybersecurity education programs to identify gaps in curriculum, teaching methods, and accessibility. Research should focus on ensuring that professionals are equipped with the skills needed to manage AI systems responsibly and that the general public understands the broader societal impacts of AI in cybersecurity.

*2) Future Direction:*

AI-powered training simulations: Develop immersive, AI-powered training simulations for cybersecurity professionals. These simulations could adapt to individual learning needs and emerging cyber threats, offering a dynamic learning environment that prepares workers for AI-augmented cybersecurity challenges.

*H. Privacy-Preserving AI in Cybersecurity*

Balancing the need for comprehensive data analysis with privacy concerns is a persistent challenge in AI-powered cybersecurity. Future research should explore advanced privacy-preserving techniques, such as federated learning and homomorphic encryption, to enable AI systems to operate effectively without compromising users' privacy.

*1) Research Opportunity:*

Privacy-preserving techniques in cybersecurity: Investigate the application of privacy-preserving AI techniques specifically for cybersecurity. This includes exploring methods like federated learning, which allows for collaborative data analysis without centralizing sensitive information, and homomorphic encryption, which enables computations on encrypted data.

*2) Future Direction:*

Frameworks for privacy-by-design: Develop AI systems with privacy-by-design frameworks that ensure user data is protected throughout the cybersecurity process. Such frameworks should be integrated into the development of AI systems from the outset, ensuring that privacy concerns are addressed proactively rather than retroactively.

*I. Bias Mitigation in AI-Powered Cybersecurity Systems*

As bias in AI systems becomes an increasing concern, especially in high-stakes domains like cybersecurity, future research must focus on developing methods to identify, quantify, and mitigate bias in AI-powered cybersecurity tools. Bias can skew threat detection, risk scoring, and user profiling, leading to unfair outcomes.

*1) Research Opportunity:*

Bias detection and mitigation: Research new methodologies for identifying and mitigating bias in AI systems used for threat detection, user behavior analysis, and incident response. Bias mitigation techniques should be adaptive, continuously monitoring AI systems for biased outcomes and adjusting algorithms to ensure fairness.

*2) Future Direction:*

Adaptive bias mitigation tools: Develop tools that can automatically detect and address biases in real-time, ensuring that AI systems are fair and equitable across diverse user populations and evolving cybersecurity environments.

## VI. CONCLUSION

The rapid integration of artificial intelligence (AI) into cybersecurity marks a significant turning point, offering unparalleled advancements in threat detection, response, and risk management. However, this progress brings with it complex ethical, regulatory, and operational challenges that must be carefully addressed. Throughout this paper, we have examined the multifaceted regulatory landscape, explored the ethical imperatives, and highlighted innovative AI applications that are reshaping the future of cybersecurity.

The historical evolution of AI regulation reveals that early efforts were largely theoretical, with more concrete frameworks emerging only in recent years. Despite notable progress, current regulatory approaches remain fragmented, and the rapid pace of AI advancements continues to outstrip governance mechanisms. To address these gaps, we propose the development of adaptive regulatory frameworks capable of evolving in real-time to meet new ethical and technological challenges, ensuring that AI remains both innovative and secure.

Ethical considerations, including fairness, transparency, accountability, and privacy, are critical for the responsible deployment of AI in cybersecurity. As AI systems become more autonomous, the need for explainability and human oversight becomes increasingly urgent. Ensuring that AI systems operate without bias, respect privacy, and maintain accountability is essential for building public trust and safeguarding against unintended harm.

Looking ahead, future research must focus on developing AI systems that not only enhance cybersecurity capabilities but also adhere to ethical standards and regulatory frameworks. Collaborative efforts across borders, industries, and research institutions are essential to ensure that AI-powered cybersecurity systems are globally aligned, adaptable, and resilient against emerging threats. Additionally, the continued promotion of AI literacy will be vital to preparing both cybersecurity professionals and the general public to engage with these technologies responsibly.

In summary, the future of AI in cybersecurity will be defined by our ability to navigate the balance between innovation and regulation, ethics and efficiency, and autonomy and oversight. By fostering a proactive, globally coordinated approach to AI governance, we can harness the full potential of AI to protect digital infrastructures while maintaining ethical integrity and societal trust.